\documentclass[twocolumn,aps,showpacs]{revtex4}
\usepackage{amsmath}
\usepackage{amssymb}
\usepackage{color}
\usepackage[dvips]{graphicx}

\begin{document}

\title{Correlations in orbital angular momentum of spatially entangled paired photons generated in parametric
downconversion}

\author{Clara I. Osorio$^1$, Gabriel Molina-Terriza$^{1,2}$ and Juan P. Torres$^{1,3}$}

\affiliation{$^{1}$ICFO-Institut de Ciencies Fotoniques,
Mediterranean Technology Park, 08860 Castelldefels (Barcelona)
Spain}

\affiliation{${^2}$ICREA-Institucio Catalana de Recerca i Estudis
Avancats, 08010 Barcelona, Spain}

\affiliation{$^{3}$ Dept. Signal Theory and Communications,
Universitat Politecnica de Catalunya, Campus Nord, 08034 Barcelona
Spain}

\email{juan.perez@icfo.es}

\begin{abstract}
What are the orbital angular momentum correlations between
spatially entangled photon pairs generated in spontaneous
parametric down-conversion?  We show that the answer to this
question can be given in two alternative, although complementary,
ways. The answer posed in this letter explains satisfactorily the
{\em seemingly contradictory} results obtained in different
experiments, and theoretical approaches.
\end{abstract}


\maketitle

Within the paraxial quantum optics regime, the orbital angular
momentum (OAM) provides an useful description of the spatial
degree of freedom of photons. Photons whose spatial waveform
contains an azimuthal phase dependence of the form $\sim \exp
\left( i m \varphi\right)$, carry an OAM of $m\hbar$ per photon
\cite{allen1}. Photons with diverse spatial waveforms can be
easily generated, detected, and controlled. Therefore, the OAM
offers a physical resource where to explore deeper quantum
features not present in the $2$-dimensional Hilbert space
addressed when using the polarization \cite{mair1,molina1}.
Indeed, it allows to readily tailor the number of effective
dimensions of the Hilbert space \cite{torres1}.

During the last few years, several quantum features based on the
capacity of the OAM of photons to go beyond a $2$-D Hilbert space
have been demonstrated (see \cite{nature1} and references inside)
using spontaneous parametric down-conversion (SPDC). These include
the demonstration of the violation of bipartite, three dimensional
Bell inequalities \cite{vaziri1}, the implementation of the so
called {\it quantum coin tossing} protocol with qutrits
\cite{molina2}, and the generation of quantum states in ultra-high
dimensional spaces \cite{barreiro1}.

All of these experiments make use of the existence of specific
quantum OAM correlations between the two entangled photons
generated in the SPDC process. Several experiments
\cite{mair1,walborn1,molina4,white1} seems to support the validity
of the selection rule $m_p=m_1+m_2$, where $m_p \hbar$ is the OAM
per photon of the classical pump beam, and $m_1$ and $m_2$ are the
winding numbers of the modes into which the quantum state of the
signal and idler photons are projected, respectively. Some other
experiments \cite{burnham1,barbosa2,molina3}, while not directly
measuring the OAM of the downconverted photons, demonstrate the
existence of ellipticity of the spatial waveform , which should
make possible the detection of photons with $m_p \ne m_1+m_2$.
Under some restrictive conditions, the selection rule
$m_p=m_1+m_2$ can be derived from first principles
\cite{barbosa1,franke1,torres1}, although, as it will be shown
below, the same rule address different physical quantities. The
presence of Poynting vector walk-off can also strongly modify OAM
correlations \cite{torres2}.

All this raises the question of what are the OAM correlations
between the downconverted photons generated in SPDC, i.e., under
which conditions the OAM of the entangled photons fulfill the
selection rule $m_p=m_1+m_2$. Here we show that this question can
be formulated in two complementary scenarios, so that in each
scenario the sough-after OAM correlations can be different. The
existence of previous {\em apparently contradictory} results is
due to the fact that the sought-after OAM correlations are
different.

In one scenario, the spatial properties of all the pairs of
photons generated are considered, therefore the {\em global} mode
function is obtained adding coherently all such possibilities. In
another scenario, which is relevant for current experimental
applications, a small section of the full down-conversion cone is
considered. Only certain probability amplitudes are now
considered. Under these conditions, the non-collinear SPDC
geometry and the presence of spatial walk-off can greatly modify
the OAM correlations observed.

We consider a nonlinear crystal of length $L$, illuminated by a
monochromatic laser pump beam propagating in the $z$ direction,
with frequency $\omega_p$. The spatial shape of the pump beam at
the input face of the nonlinear crystal ($z=0$), in the transverse
wavevector domain, writes $E_p^{+} \left( {\bf \bar p} \right)=E_0
\left( {\bar p}_x+i {\bar p}_y \right)^{m_p} \exp \left( -|{\bf
\bar p}|^2 w_0^2/4 \right)$, which corresponds to a beam which
carries an OAM of $m_p\hbar$ per photon. $E_0$ is a normalizing
constant, ${\bf \bar p}=({\bar p}_x,{\bar p}_y)$ is the transverse
wavevector and $w_0$ is the beam width. The signal and idler
photons are assumed to be monochromatic, with
$\omega_s=\omega_i=\omega_p/2$, where $\omega_{s,i}$ are the
frequencies of the signal and idler photons. This is justified by
the use of narrow-band interference filters in front of the
detectors.

The photons are known to be generated into cones, whose shape is
determined by the phase matching conditions inside the crystal.
For the sake of clarity, let us consider first noncritical, i.e.
negligible walk-off, non-collinear SPDC in a periodically poled
nonlinear crystal. The angle of the down conversion cone is
assumed to be small, so that the polarization \cite{migdall1} and
refractive index do not show noticeable changes with the direction
of propagation. Similarly, the nonlinear coefficient is assumed to
be constant along the down-conversion cone.

The SPDC process can be described in the interaction picture by an
effective Hamiltonian given by \cite{klyshko1} $H_I=\epsilon_0
\int_{V} dV \,\chi^{(2)} E_p^{+} E_s^{-} E_i^{-} + \mbox{c.c}$,
where $E_{s}^{-} \left( {\bf x},z,t \right)\propto \int dK_s d{\bf
P} \exp \left( -i {\bf P} \cdot {\bf x}-iK_{s} z +i\omega_{s}t
\right) a^{\dag} \left(K_s,{\bf P} \right)$ refers to the
negative-frequency part of the signal electric field operators.
Similarly for the idler photon. The two-photon quantum state
$|\Psi\rangle$ at the output face of the nonlinear crystal, within
the first order perturbation theory, can be written as
$|\Psi\rangle=\int d {\bf P} d {\bf Q} \,\Phi \left( {\bf P},{\bf
Q} \right) a_s^{\dag} \left({\bf P} \right) a_i^{\dag} \left({\bf
Q} \right)|0, 0 \rangle$, where $\bf P$ and $\bf Q$ are the
transverse wavevector for the signal and the idler respectively,
and the mode function $\Phi$ is given by
\begin{eqnarray}
\label{ModeFunction}
    & & \Phi \left({\bf P},{\bf Q}
    \right)= E_p \left( {\bf P}+{\bf Q} \right)  \mathrm{sinc}
    \left( \frac{\Delta_k L}{2} \right) \nonumber \\
     & & \exp \left\{ i \frac{\Delta_k L}{2}+ i \left[ K_s \left( {\bf P} \right)+ K_i \left( {\bf Q} \right) \right]
     L
    \right\}
\end{eqnarray}
where $\Delta_k=K_p \left( {\bf P}+{\bf Q} \right)-K_s \left( {\bf
P} \right)-K_i \left( {\bf Q} \right)$, the wavevectors
($j=s,i,p$) write $K_j \left( {\bf P} \right)=\left[
\left(\omega_j n_j/ c \right)^2-|{\bf P}|^2 \right]^{1/2}$,
depends on the modulus of the corresponding transverse
wavevectors, and $n_j$ are the corresponding refractive index. The
mode function of the biphoton in the spatial domain (${\bf
x}_1,{\bf x}_2$) is the spatial Fourier transform of the mode
function given by Eq. (\ref{ModeFunction}).

We can write $|{\bf P}+{\bf Q}|^2=\rho_s^2+\rho_i^2+2 \rho_s
\rho_i \cos \left(\varphi_s-\varphi_i \right)$, where
$\rho_{s}=|{\bf P}|$, and $\varphi_s=\tan^{-1} P_y/P_x$ are the
modulus and phase of the transverse wave vector ${\bf P}$ in
cylindrical coordinates. For the idler photon we have, similarly,
$\rho_{i}=|{\bf Q}|$ and $\varphi_i=\tan^{-1} Q_y/Q_x$. One can
write $\mathrm{sinc} \left( \Delta_k L/2 \right) \exp \left( i
\Delta_k L/2 \right)=\sum_{l=-\infty}^{\infty} {\cal H}_l \left(
\rho_s,\rho_i \right) \exp \left\{ i l \left( \varphi_s-\varphi_i
\right) \right\}$. $K_{s,i}$ depends on the moduli $\rho_{s,i}$,
respectively. The pump beam can also be written as
\begin{eqnarray}
& & E_p =E_0 \exp \left\{ -\frac{ \left[\rho_s^{2}
+\rho_i^{2}+\rho_s \rho_i \cos \left(\varphi_s-\varphi_i \right)
\right] w_0^{2}}{4}\right\}
 \nonumber
\\
& & \times \sum_{l=0}^{m_p}  \left( \begin{array}{c} m_p \\
l
\\\end{array} \right)  \rho_s^l \rho_i^{m_p-l} \exp \left\{ i l \varphi_s+i
\left( m_p-l \right) \varphi_i \right\}
\end{eqnarray}
Therefore, the mode function given by Eq. (\ref{ModeFunction}) can
be written as
\begin{equation}
 \label{Modefunction2}
 \Phi \left( {\bf P},{\bf Q}
    \right)=\sum_{m=-\infty}^{\infty} {\cal G}_m \left(\rho_s,\rho_i \right) \exp \left[ i m \varphi_s
    +i \left( m_p-m \right) \varphi_i \right]
\end{equation}

The main conclusion to be drawn from Eq. (\ref{Modefunction2}) is
that, if polarization, refractive index and nonlinear coefficient
show negligible azimuthal variations around the down-conversion
cone, the OAM correlations of the spatial waveform of the biphoton
state fulfill $m_p=m_1+m_2$ \cite{barbosa1}. Importantly, this
result requires considering {\em the whole spatial waveform of the
downconverted photons, i.e. the full down-conversion cone}.
Notwithstanding, these {\em are not} the OAM correlations that
typical quantum information experiments based on spatial
entanglement measure. An experiment aimed at detecting the {\em
global} OAM of the downconverted photons is a significant
experimental challenge that it is yet to be solved.

\begin{figure}
\centering\includegraphics[width=1\columnwidth]{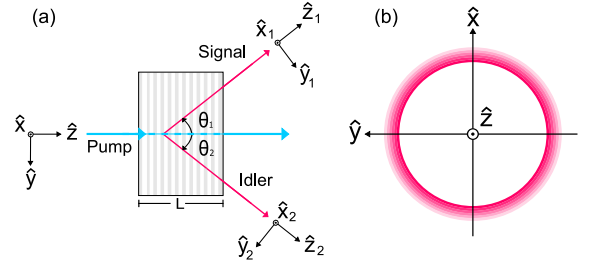}
\caption{Schematic diagram of a non-collinear SPDC. (a) Top view
of the non-collinear configuration. (b) The down-conversion cone.}
\end{figure}

All relevant experiments reported to date detect only a small
section of the full down conversion cone. In other words, the wave
vectors of the signal and idler photons belong to a narrow bundle
around the corresponding central wavevectors, i.e. ${\bf P}={\bf
P}_0+\Delta {\bf P}$ and ${\bf Q}={\bf Q}_0+\Delta {\bf Q}$. As
shown in Fig. 1, the signal photon propagate along the direction
${\hat z}_{1}$ with longitudinal wavevector $k_s \left( {\bf p}
\right) =\left[ \left( w_s n_s/c \right)2-|{\bf p}|^2
\right]^{1/2}$, and transverse wavevector ${\bf p}=(p_x,p_y)$, so
that $\Delta P_x=p_x$ and $\Delta P_y=\cos \theta_1  p_y-\sin
\theta_1 k_s$. And similarly for the idler photon, which
propagates in the direction ${\hat z}_2$ with longitudinal
wavevector $k_i \left( {\bf q} \right)$, and transverse wavevector
${\bf q}$, so that $\Delta Q_x=q_x$ and $\Delta Q_y=\cos \theta_2
q_y-\sin \theta_2 k_i$. We restrict ourselves to the case
$\theta_1=-\theta_2=\theta$.

The quantum state of the biphoton at $z_1=L/\cos \theta$ can be
written as $|\psi\rangle=\int d{\bf p} d {\bf q} \,\Phi \left(
{\bf p},{\bf q} \right) a_s^{+} \left( {\bf p} \right) a_i^{+}
\left( {\bf q} \right)|0, 0 \rangle$, where the mode function
writes \cite{torres3}
\begin{eqnarray}
\label{delta} & & \Phi \left( {\bf p},{\bf q} \right)=E_p \left(
p_x+q_x, \delta_0 \right)  \mathrm{sinc}
    \left( \frac{\delta_k L}{2} \right) \nonumber \\
    &  & \exp \left\{ i \frac{\delta_k L}{2}+i \left[ k_s \left( {\bf p} \right) +k_i
    \left( {\bf q} \right) \right] \frac{L}{\cos \theta}
    \right\}
\end{eqnarray}
where $\delta_k=k_p- \left( k_s +k_i \right) \cos
\theta-(p_y-q_y)\sin \theta$ and $\delta_0=\left( p_y+q_y \right)
\cos \theta-(k_s+k_i)\sin \theta$.

The mode function given by Eq. (\ref{delta}) shows ellipticity in
the (${\bf p},{\bf q}$) domain, as has been demonstrated
experimentally \cite{barbosa2,molina3}. An increasing degree of
ellipticity of the spatial mode function enhances the quantum
probability amplitude of paired photons with $m_p \ne m_1+m_2$. To
get further insight in the nature of the OAM correlations, let us
consider that the idler photon is projected into a gaussian mode
($m_2=0$), so that the quantum state of the signal photon is
described by the reduced mode function $\Phi_s \left( {\bf p}
\right) \propto \int d{\bf q} \Phi \left( {\bf p},{\bf q} \right)
\exp \left( -|{\bf q}|^2 w_1^2/4 \right)$, where $w_1$ is the beam
width of the idler mode. To elucidate the OAM content of the
signal photon, one has to project the spatial mode function into
spiral harmonics $\exp \left( im \varphi \right)$. The weights for
each $m$ of such decomposition gives us the sought-after OAM
decomposition \cite{molina1}.

\begin{figure}
\centering\includegraphics[width=0.75\columnwidth]{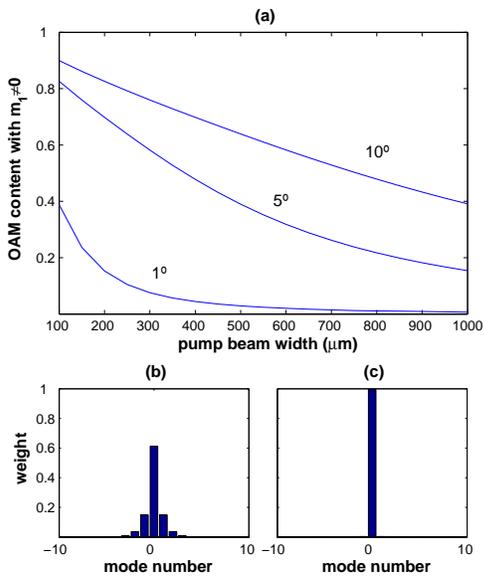}
\caption{Orbital angular momentum of the signal photon in
non-collinear SPDC in a $L=10$ mm long PPKTP crystal. The pump
beam is gaussian ($m_p=0$). Both photons traverse $2$-f systems.
The idler photon is detected with $\bf q=0$. (a) Total weight of
the OAM modes with $m_1 \ne 0$, as a function of the pump beam
width ($w_0$). The label designate the non-collinear angle. (b)
and (c) OAM mode distribution for $\varphi=1^{\circ}$ for two
values of the pump beam width: $w_0=100 \mu$m, and $w_0=1000
\mu$m.}
\end{figure}

Fig. 2(a) shows the total weight corresponding to OAM modes with
$m_1 \ne 0$, which is a measure of the degree of violation of the
selection rule $m_p=m_1+m_2$, for different values of the
non-collinear angle and the pump beam width. Notice that the
larger the non-collinear angle $\theta$, and the smaller the pump
beam width $w_0$, the larger is the probability to detect signal
photons with $m_1 \ne 0$. Figs. 2(b) and 2(c) show two OAM
distributions for $\theta=1^{\circ}$, one which shows an OAM
distribution with modes with $m_1 \ne 0$, while the other shows a
single peak for $m_1=0$.

The strength of the violation of the selection rule $m_p=m_1+m_2$
can be quantified through the non-collinear length \cite{torres2}
$L_{nc}=w_0/\sin \varphi$. If the crystal length is much smaller
than the non-collinear length ($L \ll L_{nc}$), the ellipticity of
the mode function is small, and thus the selection rule
$m_p=m_1+m_2$ is fulfilled. This turns out to be the case of
nearly all of the experiments that make use of the OAM of photons
\cite{mair1,vaziri1,molina2,barreiro1,walborn1,white1,molina4}.
Typical non-collinear angles and crystal lengths used in these
experiments are $\theta \simeq 1-2^{\circ}$ and $L \simeq 1-5$ mm.
For $w_0=500 \mu$m, one obtains $L_{nc} \simeq 15-30$ mm, so that
$L < L_{nc}$. On the contrary, if $L \ge L_{nc}$, strong
departures from the selection rule $m_p=m_1+m_2$ are expected.
This is the experimental configuration in
\cite{burnham1,barbosa2,molina3}, due to the use of a highly
focused pump beam or longer crystals. For $w_0=90 \mu$m and
$\theta=2^{\circ}$, one has $L_{nc} \simeq 2.5$ mm.

According to Eq. (\ref{Modefunction2}), the OAM of the spatially
entangled photons fulfill the relationship $m_p=m_1+m_2$, while
Fig. 2 shows that strong departures from this selection rule can
be observed if highly focused pump beams, highly non-collinear
configurations or longer nonlinear crystals are used. Actually,
{\em we are describing the same quantum process in two
complementary scenarios}. Fig. 2 would give us the OAM
correlations measured in typical experiments that use the OAM as
physical resource for quantum information, thus it is relevant for
experimental configurations currently used. In this scenario, the
fulfillment of the condition $m_p=m_1+m_2$ depends on the pump
beam width and the non-collinear angle, as dictated by the
interplay between the non-collinear and crystal lengths.

On the other hand, Eq. (\ref{Modefunction2}) corresponds to a
global view of the SPDC process, where the full down-conversion
cone is considered. The question of angular momentum conservation
balance in SPDC requires the simultaneous consideration of the
angular momentum of the electronic spins and orbitals, the
crystalline structure of the nonlinear crystal and of the
electromagnetic field \cite{bloembergen1}. The analysis presented
here might be an important step towards clarifying how angular
momentum is effectively conserved, since to evaluate conservation
laws, one should take into account {\em all probabilities
amplitudes} that contribute to the quantum process.

Another important effect that might modify the OAM correlations is
the presence of Poynting vector walk-off in some, or all, of the
interacting waves. In type I and type II SPDC configurations, some
of the interacting waves are extraordinary waves, thus show
Poynting vector walk-off. For the sake of simplicity, let us
consider a type I configuration, where only the pump beam presents
spatial walk-off. An initially gaussian pump beam, at each
position $z$ inside the nonlinear crystal can be decomposed into
spiral harmonics as
\begin{eqnarray}
& & E_p \left( {\bf \rho_p}, z  \right)=E_0 \exp \left\{ -
\rho_p^2 \left( \frac{w_0^2}{4}+ i\frac{z}{2k_p^0}\right)\right\}
\nonumber \\
& & \times \sum_{n=-\infty}^{\infty} J_{n} \left( z \rho_p \tan
\rho_0 \right) \exp \left \{i n \varphi_p \right\}
\end{eqnarray}
where $\rho_0$ is the spatial walk-off angle, $\varphi_p$ the
azimuthal angle in cylindrical coordinates and $J_n$ are Bessel
functions of the first kind, and $k_p^0$ is the longitudinal
wavevector. As shown in Fig. 3, the OAM distribution of the pump
beam increases its width with distance $z$.  A gaussian beam
incident from air into the nonlinear crystal is no longer a
gaussian beam, i.e. the pump beam is no longer a wave with
$m_p=0$.

This effect becomes noticeable only if $L
> L_w$, where the walk length writes $L_w=w_0/\tan\rho_0$. Fig.
3(a) shows the initial OAM distribution of a gaussian beam, which
corresponds to a single peak with $m_p=0$. Figs 3(b) and 3(c) show
the OAM distribution at $L=5$ mm for a walk-off angle of
$\rho_0=5^{\circ}$. For highly focused pump beams, the walk off
length can become smaller than the crystal length, thus the OAM
distribution no longer shows a single peak, as shown in Fig. 3(b).
In this case, $L_w \simeq 1$ mm. The walk off effect might be
negligible for larger beam widths, as one can see in Fig. 3(c).
Now, one obtains $L_w \simeq 12$ mm.

In the walking SPDC consider, the spatial waveform of the
two-photon can be written as
\begin{eqnarray}
\label{ModeFunction5}
    & & \Phi \left({\bf P},{\bf Q}
    \right)= E_p \left( {\bf P}+{\bf Q}\right)  \mathrm{sinc}
    \left( \Delta_k L/2 \right) \nonumber \\
      & & \exp \left\{ i \frac{\Delta _k L}{2}+ i \left[ K_s \left( {\bf P} \right)+ K_i \left( {\bf Q} \right)
     \right] L
    \right\}
\end{eqnarray}
where $\Delta_k=k_p \left( {\bf P}+{\bf Q} \right)+ \left(P_x+Q_x
\right) \tan \rho_0-k_s \left( {\bf P} \right)-k_i \left( {\bf Q}
\right)$. The phase matching function no longer depends on
$\varphi_s-\varphi_i$ due to the presence of the spatial walk-off.

The mode function of the two-photon state is a coherent sum of all
the partial amplitudes due to the possible generation of a pair of
photons at position $z$ inside the crystal \cite{rubin1}. Since
the beam moves in the $xz$ plane with angle $\rho_0$ when
propagating along $z$, the amplitude for each $z$ shows increasing
ellipticity, contributing differently to the total OAM
decomposition of the spatial waveform of the two-photon state.
\begin{figure}
\centering\includegraphics[width=1\columnwidth]{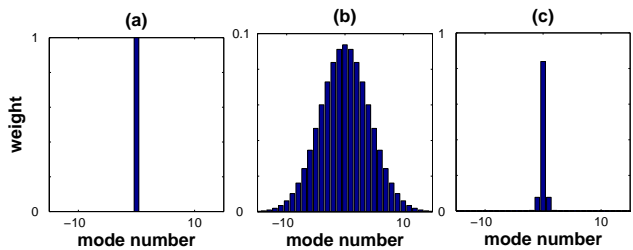}
\caption{Orbital angular momentum distribution of the pump beam in
different positions inside the nonlinear crystal. (a) $z=0$; (b)
$z=5$ mm and $w_0=100 \mu$m and (c) $z=5$ mm and $w_0=1$ mm. The
walk-off angle is $\rho_0=5^{\circ}$.}
\end{figure}

In conclusion, we have shown that the elucidation of the OAM
correlations of entangled photons generated in SPDC can be
addressed in two complementary scenarios, giving correspondingly
different OAM correlations between the photons. One scenario
considers the quantum amplitudes in the whole down-conversion
cone, while the other scenario, experimentally relevant for
quantum information protocols that make use of the OAM, pays
attention to a small section of the down-conversion cone.

The results obtained here are of great importance for other
configurations where entangled paired photons are generated. This
is the case of pairs of photons generated through two-photon Raman
transitions in electromagnetic induced transparency schemes
\cite{harris1}. Recently, the OAM correlations between stokes and
anti-stokes photons has been measured in trapped Rubidium cold
atoms in a counter-propagating nearly collinear geometry
\cite{inoue1}. The measured correlations comply with the selection
rule $m_p-m_c=m_1-m_2$, where $m_c$ refers to the OAM per photon
of a laser control beam that counter propagates with the pump
beam. The corresponding selection rule for counter propagating
signal and idler photons \cite{torres3} in SPDC would be
$m_p=m_1-m_2$. We expect that a similar experiment in a highly
noncollinear geometry, such a transverse emitting configuration
\cite{harris3}, would yield strong departures from the selection
rule $m_p-m_c=m_1-m_2$ measured for a nearly collinear
geometry.

This work was supported by projects FIS2004-03556 and
Consolider-Ingenio 2010 QOIT from Spain, by the European
Commission (Qubit Applications (QAP), Contract No. 015848) and by
the Generalitat de Catalunya.

\end{document}